\documentstyle[pra,aps]{revtex}
\begin{document}
\twocolumn[\hsize\textwidth\columnwidth\hsize\csname @twocolumnfalse\endcsname
\draft
\author{E. Solano,$^{1,2}$ P. Milman,$^{1}$ R. L. de Matos Filho,$^{1}$ and N. Zagury
$^{1}$}
\title{Manipulating motional states by selective vibronic interaction in two
trapped ions}
\address{$^{1}$Instituto de F\'{\i}sica, Universidade Federal do Rio de Janeiro,
Caixa Postal
68528, 21945-970 Rio de Janeiro, RJ, Brazil \\
$^{2}$Secci\'{o}n
F\'{\i}sica, Departamento de Ciencias, Pontificia Universidad
Cat\'{o}lica del Per\'{u}, Apartado 1761, Lima, Peru}
\date{\today}
\maketitle

\begin{abstract}
We present a selective vibronic interaction for manipulating motional states
in two trapped ions, acting resonantly on a previously chosen vibronic
subspace and dispersively on all others. This is done respecting technical
limitations on ionic laser individual addressing. We discuss the generation of
Fock states and entanglement in the ionic collective motional degrees of
freedom, among other applications.
\end{abstract}

]

\vskip2pc

Coherent manipulation of vibronic states of two or more trapped
ions has become a subject of increasing interest in the last few years\cite{review}. A
number of experimental and theoretical advances has been done recently in
this field, aiming to make applications possible and to test basic features
of quantum mechanics. In this context, Bell states and multiparticle
entanglement of the ionic internal degrees of freedom have been studied \cite
{sorensen2,SMZ,sorensen3} and partially realized \cite{turchette,sackett}.
Beside their fundamental importance, such states are essential ingredients
for the implementation of several interesting applications, as quantum logic
and quantum computing using trapped ions\cite{review,cirac,sorensen1}. One
of the main problems in realizing some of the above proposals is the
technical difficulty in the individual ionic laser illumination. Whereas
many efforts are being done to overcome this problem \cite{leibfried,nagerl}
, it still represents an obstacle to the existence of realistic proposals
for the deterministic manipulation of the motional degrees of freedom of two
or more trapped ions.

Recently, M\o lmer and S\o rensen\cite{sorensen2} have shown that if one
illuminates two two-level ions with two lasers that act dispersively in the
blue and red first side band transitions, it is possible to find an
effective interaction, in the Lamb-Dicke limit, that causes a direct
transition between the ground and the doubly excited electronic levels of
the two ion system, without changing their vibrational state. This permits
the deterministic generation of entangled states that are linear
combinations of these two electronic states. In ref.\cite{SMZ}, the
combination of this interaction with a carrier transition permits the
generation of all electronic Bell states. This leads to more general
applications, as teleportation and entanglement swapping\cite{SLMZ}. All
these proposals are mostly concerned with the deterministic manipulation of
internal states of the ions. However, it is also important to find ways for
manipulating the vibrational degrees of freedom of two or more trapped ions,
respecting the limitations imposed on their individual addressing.

In this article we present an effective interaction consisting of two
dispersive Raman pulses simultaneously illuminating two trapped ions, which
opens the possibility to manipulate coherently their vibrational motion. As
the electronic Stark shifts, induced by the Raman pulses, depend on the
motional state of the ions, the resulting dynamics can be described by a
Jaynes-Cummings-like interaction acting distinctly on different subspaces
of their vibronic Hilbert space. As we will show below, the frequencies
of the two dispersive Raman pulses may be chosen in such a way that the
effective interaction becomes resonant to a previously chosen vibronic
subspace while remaining non-resonant to others. This enables us to excite
selectively a desired subspace inside the motional Hilbert space of the
ions. Vibrational state engineering of both center of mass and relative
motion can then be done by means of this special property.

We consider two two-level ions of mass $m$, confined to move in the $z$
direction in a Paul trap. They are cooled down to very low temperatures \cite
{king} and may perform small oscillations around their equilibrium
positions, $z_{10}=-d/2,$ $z_{20}=d/2.$ We denote by $\hat{Z}=(\hat{z}_1+
\hat{z}_2)/2$ and $\hat{z}=(\hat{z}_1-\hat{z}_2)/2$ the center of mass and
relative position operators, respectively. Both ions are simultaneously
illuminated by two classical homogeneous Raman effective pulses $\vec{E}_{I}=
\vec{E}_{0I}e^{i(\vec{q}_{1}\cdot\vec{r}-\omega _{I}t)}$ and $\vec{E}_{II} =
\vec{E}_{0II}\,e^{i(\vec{q}_{2}\cdot\vec{r}-\omega _{II}t)},$ with wave
vectors $\vec{q}_1=\vec{q}_{2}=\vec{q},$ parallel to the $z$ direction and
angular frequencies $\omega_{I}$ and $\omega_{II}$. The Raman pulses
frequencies will be chosen to be quasi-resonant with a long-living
electronic transition between two ionic hyperfine levels $|\uparrow_{j}
\rangle $ and $|\downarrow_{j} \rangle $ (j=1,2), with energy $\hbar\omega_0$
and $0$, respectively. The total Hamiltonian of the system may be written,
in the optical RWA approximation, as
\begin{equation}
\hat{H}=\hat{H}_0+\hat{H}_{{\rm int}},
\end{equation}
with
\begin{eqnarray}
\hat{H}_0&=&\hbar \omega_{o}(\hat{S}_{+1}\hat{S}_{-1}+\hat{S}_{+2}\hat{S}
_{-2}) +\hbar \nu \hat{a}^\dagger \hat{a} + \hbar \nu_r \hat{b}^\dagger \hat{
b} ,  \nonumber \\
\hat{H}_{{\rm int}}&=&\hbar \Omega\, e^{iq\hat{Z}} \left( \hat{S}_{+1}\,e^{iq
\hat{z}/2} +\hat{S}_{+2}\,e^{-iq\hat{z}/2}\right) \times  \nonumber \\
&&\left(e^{-i\omega_{I}t}+e^{-i\omega_{II}t}\right)+{\rm H.c.}
\end{eqnarray}
Here $\widehat{S}_{+j}=|\!\!\uparrow_j\rangle\langle \downarrow_j\!\!|$ is
the flip operator associated with the electronic transition $
|\downarrow_j\rangle\rightarrow|\uparrow_j\rangle$ in the ion $j.$  The
operators $\hat{a}$ and $\hat{b}$ ($\hat{a}^{\dagger}$ and $\hat{b}^{\dagger }$) are the
annihilation (creation) operators associated with the center of mass mode of frequency $\nu$ and the
relative vibrational mode of frequency $\nu_r,$ respectively. For simplicity, we have assumed that the same Rabi frequency $\Omega$ (taken as real) is associated with both lasers.

We start by taking the  frequencies $\omega_I$ and $\omega_{II}$ as:
\begin{eqnarray}  \label{freq}
\omega_{I} &=& \omega_0 + k \nu +k_r {\nu}_r- \delta  \nonumber \\
\omega_{II} &=& \omega_0 + \delta,
\end{eqnarray}
with $\delta \ll \nu, {\nu}_r .$ The frequencies are chosen so that $
\omega_{I}+\omega_{II}= k\,\nu +k_r\,\nu_r + 2\omega_0 .$

Following the usual treatment for one single ion interacting with a laser
field~ \cite{vogel}, we make the
rotating-wave-approximation (RWA) with respect to the vibrational
frequencies and select the terms that oscillate with minimum frequency. In
the interaction picture we obtain, for the interaction Hamiltonian, the
following expression:
\begin{equation}  \label{geral}
\hat{H}^{I}_{\rm int}=\hbar \Omega \left( \hat{S}^{\prime}_+\hat{a}^{\dagger k}
\hat{b}^{\dagger k_r} \hat{H}_{k,k_{r}} e^{i\delta t}+\hat{S}
^{\prime\prime}_+\hat{H}_{0,0}e^{-i\delta t}\right)+{\rm {H. c.}\,,}
\end{equation}
where
\begin{eqnarray}  \label{AandB}
\hat{S}^{\prime}_{+}&=&\hat{S}_{+1}e^{i\phi _{0}/2}+ (-)^{k_r} \hat{S}
_{+2}e^{-i\phi _{0}/2}\,,  \nonumber \\
\hat{S}^{\prime\prime}_{+}&=&\hat{S}_{+1}e^{i\phi _{0}/2}+ \hat{S}
_{+2}e^{-i\phi _{0}/2}.
\end{eqnarray}
Here $\phi _{0}=z_0d$ is the phase difference due to the equilibrium distance between the
two ions, which, in this paper, may be taken as $2n\pi$ without loss of generality , and
\begin{equation}
\hat{H}_{k,k_{r}}=(i\eta)^{k}(i\eta _{r}{})^{k_{r}}\hat{F}_{k} (\eta ^{2})
\hat{G}_{k_{r}}(\eta_{r}^{2})\,,
\end{equation}
with
\begin{eqnarray}
\hat{F}_{k}(\eta ^{2}) &=&\sum_{n} e^{-\eta ^{2}/2}\frac{n!}{(n+k)!}
L_{n}^{k}(\eta^{2})|n\rangle \langle n|\,, \\
\hat{G}_{k_{r}}(\eta _{r}^{2}) &=&\sum_{n} e^{-\eta _{r}^{2}/2} \frac{n_{r}!
}{(n_{r}+k_{r})!}L_{n_{r}}^{k_{r}}(\eta _{r}^{2})|n_{r}\rangle \langle
n_{r}|\,.
\end{eqnarray}
The parameters $\eta=q\sqrt{\hbar/4m\,\nu}$ and $\eta_{r}=q\sqrt{
  \hbar/4m\,\nu_r}$ are the Lamb-Dicke parameters corresponding to the
CM and the relative modes, respectively. $L_{n}^{m}$ are
associated Laguerre polynomials, whereas the states $|n \rangle$ and
$|n_{r} \rangle$ are the eigenvectors of the number operators
$\hat{a}^{\dagger}\hat{a}$ and $ \hat{b}^{\dagger}\hat{b}$,
respectively.

If $\delta$ is large enough so that $\delta \gg \Omega $ and for times $t$
such that $\delta t \gg 1$, it is possible to derive a two-photon effective
time-independent Hamiltonian, where both CM and relative modes can be
excited. It can be written as a sum of three terms:
\begin{equation}  \label{Hgeneral}
\hat{H}_{{\rm eff}}=\hat{H}_1+\hat{H}_2+\hat{H}_3\,,
\end{equation}
with
\begin{eqnarray}\label{H123}
&&\hat{H}_1\!=\hbar \Omega _{0}
\epsilon_{k_{r}}\hat{S}_{+1}\hat{S}_{+2}\left[ \hat{a}^{\dagger k}
\hat{b}^{\dagger k_{r}}\hat{H}_{k,k_{r}},\hat{H}
_{0,0}\right]+{\rm H. c.}   \nonumber \\
&&\hat{H}_{2}\!=\hbar\Omega_{0}(-)^{k_r}
\hat{S}_{+1}\hat{S}_{-2}\left[ \hat{a}^{\dagger k}
\hat{b}^{\dagger k_{r}}\hat{H}_{k,k_{r}},
\hat{H}_{k,k_{r}}\hat{a}^{k} \hat{b}^{k_{r}}\right]\! +{\rm {H.
c.}} \nonumber \\ &&\hat{H}_{3}\!=\frac{\hbar \Omega
_{0}}{2}\sum_{j=1,2} \hat{S}_{+j}\hat{S}_{-j}(\hat{a}^{ \dagger
k}\hat{b}^{\dagger
k_{r}}\hat{H}_{k,k_{r}}^{2}\hat{a}^{k}\hat{b}^{k_{r}}-\hat{H}_{0,0}^{2})
\nonumber \\ &&-\sum_{j
=1,2}\hat{S}_{-j}\hat{S}_{+j}(\hat{a}^{k}\hat{b}^{k_{r}}\hat{a}^{\dagger
k}\hat{b}^{\dagger k_{r}}\hat{H}_{k,k_{r}}^{2}-\hat{H}_{0,0}^{2})
+{\rm H.c.\, },
\end{eqnarray}
where $\Omega_0={\Omega^2}/{\delta}.$
The first term, $\hat{H}_1,$ gives rise to an anti Jaynes-Cummings dynamics, leading to a simultaneous excitation (or
de-excitation) of the electronic states of the two ions, accompanied by the
creation (or annihilation) of $k$ vibrational quanta in the CM mode and $k_r$
vibrational quanta in the relative mode. The factor $
\epsilon_{k_{r}}=(1+(-)^{k_{r}})$ prevents excitations of odd number of
quanta in the relative mode, so that the symmetry by exchange of the two
ions is maintained. The second term, $\hat{H}_2, $ generates a dynamics
where, simultaneously, one ion undergoes a transition from the ground to the
excited electronic state and the other ion makes a transition in the inverse
direction. This process is not accompanied by any excitation of the
vibrational modes. The third term, $\hat{H}_3, $ generates motional dependent dynamical energy
shifts in the electronic levels. Due to this
dependence, this term turns the processes induced by $\hat{H}_1$ and $\hat{H}
_2$ more or less resonant, depending on the particular level of
excitation of the vibrational modes. Note that the sensitivity of the
energy shifts to the vibrational state of the ions increases with
increasing values of the Lamb--Dicke parameters. For not too small values of
the these parameters, it is possible to make the interaction $\hat{H}_1$
completely resonant inside a previously chosen subspace $
\left\{|\downarrow,\downarrow,N,N_r\rangle,|\uparrow, \uparrow, N+k,
N_r+k_r\rangle\right\},$ whereas remaining largely non resonant inside
other subspaces. The same applies for $\hat{H}_2$ in a given subspace $
\left\{|\downarrow,\uparrow,M,M_r\rangle, |\uparrow, \downarrow, M,
M_r\rangle\right\}$. This may occur, for example, when the Raman lasers
frequencies, originally given by Eq.~\ref{freq}, are modified and correctly
tuned to take into account the motional dependent  energy shifts.

To describe these effects in more detail, we turn our attention to specific
cases. We first consider excitations to the first blue side band of the
center of mass mode $(k=1,k_{r}=0).$ If we start from the electronic ground
state, only $\hat{H}_{1}$ and $\hat{H}_{3}$ will be effective , and we may write:
\begin{eqnarray}\label{HCM}
&&H_{{\rm eff}}=\hbar \Omega _{0}\hat{G}_{0}^{2}\{2(i\eta
)\hat{S}_{+1}\hat{S}_{+2}[\hat{a}^{\dagger
}\hat{F}_{1},\hat{F}_{0}]+ \nonumber\\
&&\frac{1}{2}(\hat{S}_{+1}\hat{S}_{-1}+\hat{S}_{+2}\hat{S}_{-2})(\eta
^{2}\hat{a}^{\dagger }\hat{F}_{1}^{2}\hat{a}-\hat{F}_{0}^{2})
\nonumber\\
&&-\frac{1}{2}(\hat{S}_{-1}\hat{S}_{+1}+\hat{S}_{-2}\hat{S}_{+2})(\eta
^{2}\hat{a}\hat{a}^{\dagger }\hat{F}_{1}^{2}-\hat{F}_{0}^{2})+{\rm
H.c.}\}\, .
\end{eqnarray}

As can be easily seen from Eq.\ref{HCM}, the energy shifts of levels $
|\downarrow ,\downarrow ,n,n_{r}\rangle $ and $|\uparrow ,\uparrow
,n+1,n_{r}\rangle $ are given by:
\begin{eqnarray}\label{delta}
&&\Delta _{\downarrow \downarrow}^{n,n_{r}}=2\hbar \Omega _{0}g_{0}^{2}(n_{r})\left[
f_{0}^{2}(n)-{\eta }^{2}(n+1)f_{1}^{2}(n)\right]
\label{delta} \\
&&\Delta _{\uparrow\uparrow }^{n+1,n_{r}}=2\hbar \Omega _{0}g_{0}^{2}(n_{r})\left[ {{
\eta }^{2}}(n+1)f_{1}^{2}(n)-f_{0}^{2}(n+1)\right] ,
\nonumber\end{eqnarray}
respectively. In Eq.~(\ref{delta})
\begin{eqnarray}
f_{k}(m) &=&e^{-\eta ^{2}/2}\frac{m!}{(m+k)!}L_{m}^{k}(\eta ^{2})  \nonumber
\\
g_{k}(m) &=&e^{-\eta _{r}^{2}/2}\frac{m!}{(m+k)!}L_{m}^{k}(\eta _{r}^{2})\, .
\end{eqnarray}
By properly adjusting the laser frequencies we may put them in
resonance with the Stark shifted levels associated with a previously
chosen vibronic subspace \nopagebreak[4]{$\{|\downarrow ,\downarrow ,N,N_{r}\rangle$
$,|\uparrow ,\uparrow ,N+1,N_{r}\rangle \},$} while preventing resonant
transitions in other subspaces with $n\neq N$ and $n_{r}\neq N_{r}.$
This can be done by modifying the laser frequencies to
\begin{equation}\label{newfreq}
\omega _{I}=\tilde{\omega}_{0}+\nu -\delta ,\,\,\,\,\omega _{II}=\tilde{
\omega}_{0}+\delta ,
\end{equation}
where $2\tilde{\omega}_{0}=2\omega _{0}+\Delta _{\uparrow \uparrow}^{N+1,N_{r}}-\Delta
_{\downarrow\downarrow }^{N,N_{r}}$ is the renormalized splitting of the levels.
Notice that, for very small values of the Lamb-Dicke parameters, the motional
dependence of the dynamical Stark shift disappears.
For this reason it is important to work beyond the Lamb-Dicke regime in
order to effectively select a chosen subspace out of the whole vibronic
Hilbert space. It is noteworthy to mention that
for special values of the Lamb-Dicke parameter ${\eta },$ it may happen that energy shifts $
\Delta _{\uparrow\uparrow }^{n+1,n_{r}}$ and $\Delta _{\downarrow \downarrow}^{n,n_{r}}$ are equal  for
certain values of $n,$ irrespectively of the state of the relative
vibrational mode.
For example, for ${\eta \simeq 0.51,0.42,0.24,}$
transitions inside the subspaces $\{|\downarrow ,\downarrow ,N,N_{r}\rangle
,|\uparrow ,\uparrow ,N+1,N_{r}\rangle \},$ become resonant for $N=1,2,8,$
respectively. Clearly, in this case it is not necessary to correct the laser frequencies.

In order to check  this model, numerical simulations were done using the
 time dependent Hamiltonian given in Eq. (\ref{geral}). Starting from the
state $|\downarrow ,\downarrow ,0,0\rangle $ \cite{king} and selecting the
laser frequencies as in Eq.~(\ref{newfreq}), we were able to observe
complete Rabi oscillations between the states $|\downarrow ,\downarrow
,0,0\rangle $ and $|\uparrow ,\uparrow ,1,0\rangle ,$ in agreement with the
model discussed above (see Fig.~1). In particular, for a $\pi $ pulse,  the state $
|\uparrow ,\uparrow ,1,0\rangle $ is generated with $\approx 100\%$
efficiency. For the same laser frequencies, we plot the Rabi oscillations of the population
corresponding to  the initial state  $|\downarrow ,\downarrow,1,0\rangle. $
We can see that, indeed, the Rabi oscillations for this case have a very small amplitude.

Similar results may be obtained also for transitions leading to even
excitations of the relative vibrational mode only. This is done by choosing
$k=0$ in Eq. \ref{Hgeneral} and correctly tuning the lasers to take into
account  the self energy terms. We  checked numerically that it is possible, in this case, to drive
resonantly Rabi oscillations inside selected  subspaces $\{|\downarrow
,\downarrow ,N,N_{r}\rangle ,|\uparrow ,\uparrow ,N,N_{r}+k_{r}\rangle \}.$

The Hamiltonian (\ref{Hgeneral}) could be used to generate a large set of
motional states. For example, any Fock state associated with the center of
mass motion and with even excitations of the relative vibrational mode could
be obtained from the initial state $|\downarrow ,\downarrow ,0,0\rangle $
by successively applying $\pi $ pulses with different frequencies. However,
if one is interested in generating a highly excited Fock state,  this
process could take an unsatisfactory long time. A more efficient, non
unitary, way of producing such states, as well as engineering other
vibrational states, is to start from a product of the electronic ground
state and any motional state $|\psi _{{\rm vib}}\rangle .$ By selecting the
laser frequencies, we excite only a chosen vibronic transition $\{k,k_r\}$
 with a $\pi $ pulse.
Ideally, we would end up with a superposition of the two states $|\uparrow
,\uparrow ,N+k,N_{r}+k_{r}\rangle $ and  $|\downarrow
,\downarrow \rangle \otimes (|\psi _{{\rm vib}}\rangle -|N,N_{r}\rangle
\langle N,N_{r}|\psi _{\rm vib}\rangle ).$ Measurement of  the electronic
levels projects out either the Fock state $|N+k,N_{r}+k_{r}\rangle $ or the
original state with a ``hole" in the $N,N_{r}$ component.

As a numerical example we have taken the initial state to be the
product of the electronic ground state with a coherent state of the CM
motion and the vacuum of the relative motion. We start from a coherent
state \cite{coherent } with $\bar{n}=4.0$ and excite the center of
mass transition from $n=4$ to $n=5$.  A dark event in a fluorescent
measurement should leave us with an state that is close to the Fock
state with $n=5.$ In Fig.~2 we show the results for the phonon
distribution. As expected, it is possible to make a ``hole'' in the
vibrational quanta distribution ( Fig.~2a ) while creating a quasi
Fock state ( Fig.~2b) using a $\pi $ pulse.  The state with a``hole''
will be associated with the ground electronic state, while the
approximate Fock state of Fig.~2b will be associated with the excited
electronic state. As shown in Fig.~2b, small contamination occurs
around the target Fock state because  transitions to levels other
than $n=5$ are not totally suppressed. For the case studied, the
probability of finding the approximated Fock state after fluorescence
is about $30\%$.  A Fock state with $n=6,$ for example, can now be
obtained if we apply subsequently to the ions another $\pi$ pulse
resonant to the transition $|\uparrow ,\uparrow ,5,0\rangle $
$\leftrightarrow$$|\downarrow ,\downarrow ,6,0\rangle.$ If by
measuring the electronic states we find $|\downarrow ,\downarrow\rangle $ (
{\it a priori} probability of $\approx 87\%), $ we are left with a
state very close to the Fock state $|6,0\rangle.$ In this case the
fidelity is $\approx 99\%$ (See Fig.~2c). Depopulation of a region around a certain
value of $N,$ can also be achieved by proper choice of the Lamb Dicke
parameter and the detuning $\delta $, taking advantage of the quasi
resonance character of the interaction for $n's$ very close to $N.$

Another  interesting application is the generation of maximally entangled states
inside the subspace $|1,0\rangle ,|0,2\rangle ,|1,2\rangle ,|0,0\rangle $ of
the vibrational modes. Starting from the electronic and vibrational ground
state, a $\pi /2$ pulse with a given frequency would generate the state
$\frac{1}{\sqrt{2}}\left[ |\downarrow ,\downarrow ,0,0\rangle +|\uparrow
  ,\uparrow ,0,2\rangle \right] $. A subsequent $\pi $ pulse with another
frequency would then lead to the state $\frac{1}{\sqrt{2}}|\uparrow ,\uparrow
\rangle \left[ |1,0\rangle +|0,2\rangle \right] $ which is a maximally
entangled state. Similar procedures could also lead to any maximally entangled
state.

Entanglement transfer may also be achieved between the internal and
motional degrees of freedom\cite{milburn}.  For example, assume that
we start with the state $\frac{1}{\sqrt{2}}\left[ |\downarrow
  ,\downarrow \rangle+|\uparrow ,\uparrow \rangle
\right]\otimes|0,0\rangle.$ If we apply to the ions a $\pi$ pulse,
connecting the states $|\downarrow ,\downarrow
0,0\rangle\rightarrow|\uparrow ,\uparrow ,1,2\rangle$, we obtain
$|\uparrow ,\uparrow \rangle \otimes(|0,0\rangle+|1,2\rangle). $ In
this case the efficiency is very high since the state $|\uparrow
,\uparrow \rangle \otimes|0,0\rangle$ does not couple to any state in
an anti Jaynes-Cummings transition. Of course, the inverse process,
where entanglement is transfered from the motional degrees of freedom
to electronic one, is also possible.

In summary, we have engineered an interaction that, respecting limitations of
ionic individual addressing, enhances our possibilities of manipulating and
generating diverse vibronic states of two trapped ions.  This interaction acts
selectively in a previously chosen vibronic subspace, $\{|\downarrow
,\downarrow N,N_r\rangle,|\uparrow ,\uparrow ,N+k,N_r+k_r\rangle \},$
permitting, in principle, a complete transfer of populations inside this
subspace.  The applications mentioned above are only a few relevant examples
of what may be done by selectively addressing the initial vibronic states.

This work was partially supported by the Conselho Nacional de
Desenvolvimento Cient{\'\i}fico e Tecnol\'ogico (CNPq), the Programa de
Apoio a N\'ucleos de Excel\^encia (PRONEX) and the Funda\c c\~ao de Amparo
\`a Pesquisa do Estado do Rio de Janeiro (FAPERJ).

\begin{figure}[tbp]\label{Rabi}
\caption{ Center of mass first side  band blue excitation. The larger oscillations correspond to the population of the state $|\downarrow,
\downarrow, 0, 0 \rangle$. The smaller oscillations correspond to the population of the state $|\downarrow,\downarrow, 1, 0 \rangle.$ $\eta=0.5,$ $\delta=40\eta\Omega.$}
\end{figure}

\begin{figure}[tbp]
\caption{Figure 2. Vibrational distributions $P(n)$ for the center of mass motion after
a $\pi$ pulse resonant to the transition from $n=4$ to $n=5. $ The initial
state is the product of a coherent state with $\bar n=4.0$ and $
|\downarrow,\downarrow\rangle.$ (a): Distribution correlated to $
|\downarrow,\downarrow\rangle.$ (b): Distribution correlated to $
|\uparrow,\uparrow\rangle.$ (c) Distribution correlated to $
|\downarrow,\downarrow\rangle$ after the second pulse. $\eta= 0.3$ and $\delta = 40.0 \eta \Omega$. }
\end{figure}


\begin{references}
\bibitem{coherent } A coherent state of the center of mass motion could be generated, for example,  by applying to the trapping electrodes a classical rf signal of frequency $\nu .$
\end{references}
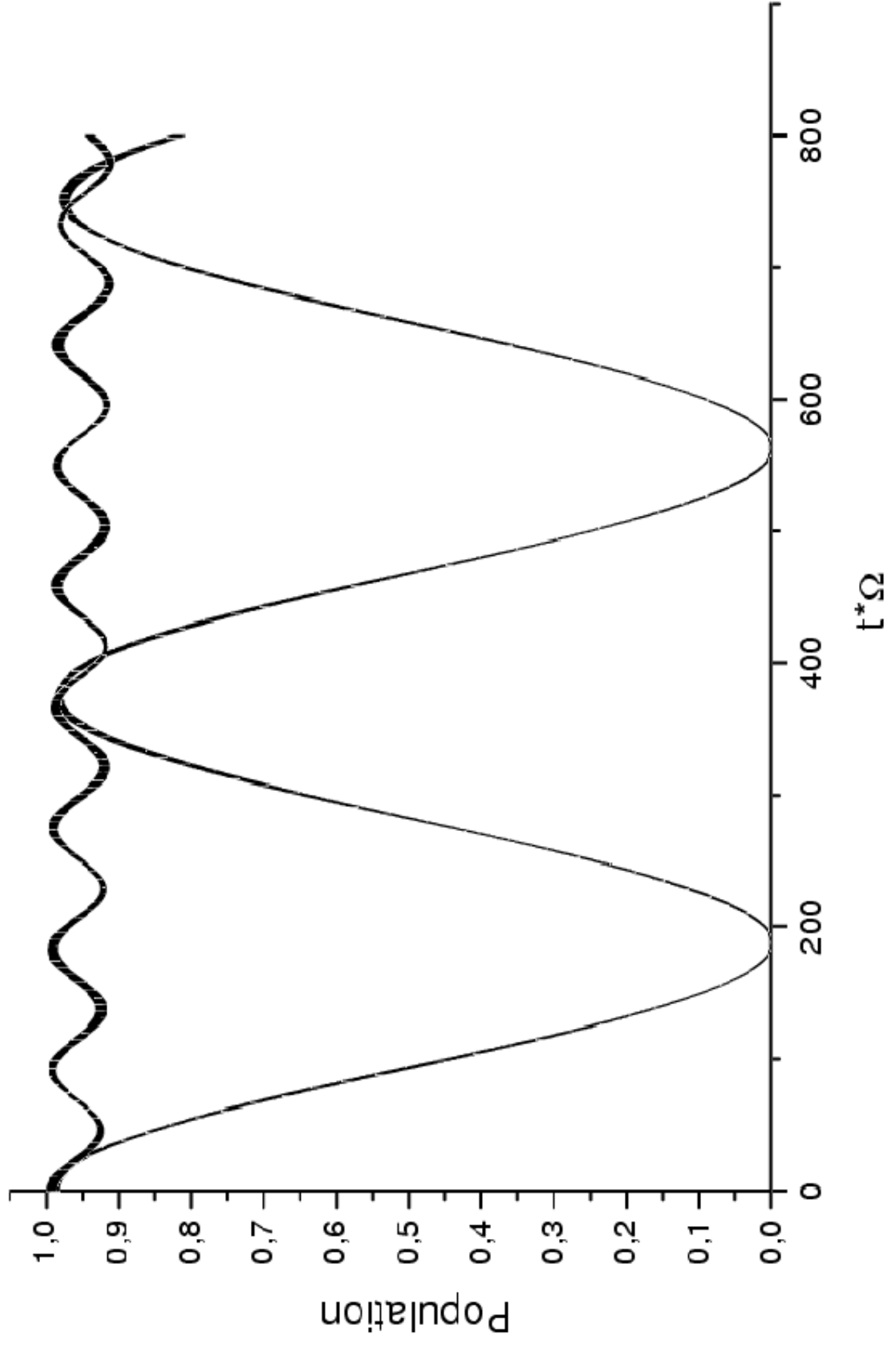
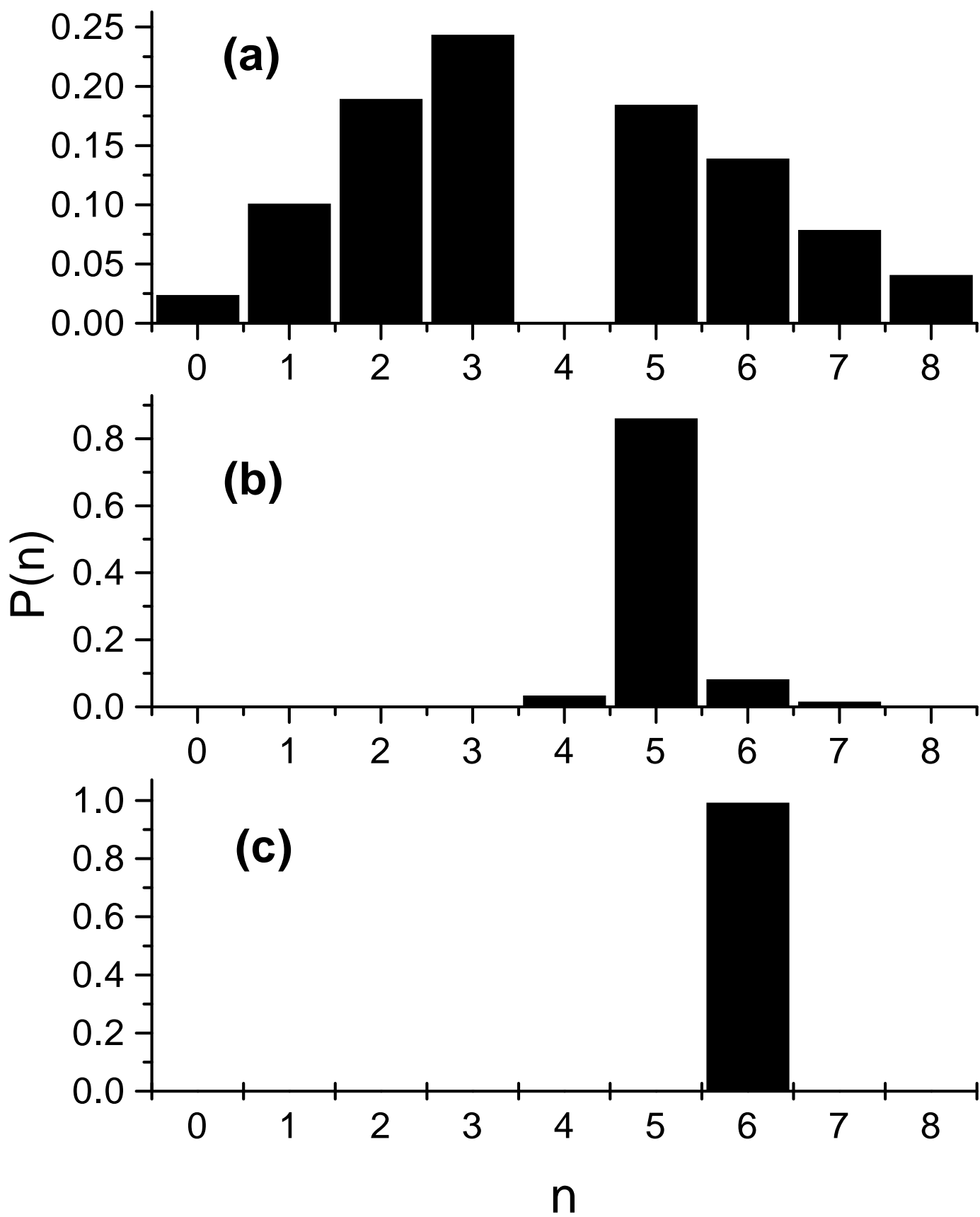
\end{document}